\documentclass[structabstract]{aa}
\usepackage{txfonts}
\usepackage{graphicx}
\usepackage{natbib}
\bibpunct{(}{)}{;}{a}{}{,}

\begin{document}
\title{Tohoku-Hiroshima-Nagoya planetary spectra library: A method for characterizing planets in the visible to near infrared}
\author{Ramsey Lundock \inst{1}
\and Takashi Ichikawa \inst{1}
\and Hirofumi Okita \inst{1}
\and Kentarou Kurita \inst{1}
\and Koji S. Kawabata \inst{2}
\and Makoto Uemura \inst{2}
\and Takuya Yamashita \inst{2}
\and Takashi Ohsugi \inst{2}
\and Shuji Sato \inst{3}
\and Masaru Kino \inst{3} }

\institute{Astronomical Institute, Tohoku University, Aoba, Sendai 980-8578, Japan
\and Astrophysical Science Center, Hiroshima University, Kagamiyama 1-3-1, Higashi-Hiroshima 739-8526, Japan
\and Department of Physics, Nagoya University, Furo-cho, Chikusa-ku, Nagoya 464-8602, Japan}

\date {Received / Accepted }

\abstract { }{There has not been a comprehensive framework for comparing spectral data from 
different planets.
  Such a framework is needed for the study of extrasolar planets and objects within the solar system. 
We have undertaken observations to compile a library of planet spectra for all planets, some moons, 
and some dwarf planets in the solar system to study their general spectroscopic and photometric natures.}
{During May and November of 2008, we acquired spectra for the planets using TRISPEC,
which is capable of simultaneous three-band spectroscopy in a wide wavelength range of 
0.45 -- 2.5 $\mu$m with low resolving power ($\lambda/\Delta\lambda \sim 140$ -- 360). 
}
{Patterns emerge from comparing the spectra. 
Analyzing their general spectroscopic and photometric natures, 
we show that it is possible to distinguish between gas planets, soil planets and ice planets.
These methods can be applied to extrasolar observations using low resolution spectrography or broad-band filters.
}
{ The present planet spectral library
is the first library to contain observational spectra for all of the solar system planets,
based on simultaneous observations in visible and near infrared wavelengths.
This library will be a useful reference for analyzing extrasolar planet spectra, and for 
calibrating planetary data sets. 
}

\keywords{ Astronomical data bases: miscellaneous - Planets and satellites: general}
\titlerunning{THN-PSL}
\authorrunning{Lundock \textit{et al.}}

\maketitle

\section{Introduction}
Over the past several decades, improved telescopes and probe missions have revealed an 
unprecedented amount of information about the planets orbiting the Sun and other stars.  
The numbers of known planets and `not quite planets' are increasing every year.
Ceres, Pluto and Eris were reclassified as dwarf planets in 2006.
Haumea and Makemake have subsequently been added to the dwarf planet category \citep{WGPSN}.
It is an open question whether small brown dwarves are distinct from large gas giants. 
Thanks to the enormous data taken so far by many authors, we know a great deal about 
the nature of each planet in the solar system.
However, the framework for discussing `planets in general' has lagged behind. 

Spectra exist for all of the solar system planets and many of the smaller objects.
In many cases to find spectral libraries it is necessary to go back to the 1970's and early 1980's 
when the first quality spectra of planets were being obtained.
Particularly prolific studies during this era were, e.g., 
McCord \& Westphal (1971), Fink et al. (1976), McCord et al. (1979), 
Fink \& Larson (1979), Clark (1980, 1981), and McCord (1981).
However, this was also a time of rapid technological development and even within a single 
research group, observation techniques would change.
Between their 1976 and 1979 observations the Fink group at Kitt Peak upgraded from a 90 inch 
telescope to a 4m telescope 
\citep{Fink76, Fink79}.
The Pollack group using the Kuiper Airborne Observatory used an uncooled circular variable 
filter wheel for their 1975 observations of Venus and Jupiter \citep{Pollack78a}.
They upgraded to a cooled filter wheel for their 1976 observations of the Galilean satellites \citep{Pollack78b}.
Few observations have been performed for spectral libraries in the past 2 decades.
Examples of more recent spectral data in broad wavelengths 
are Earth's spectra from 0.7 to 2.4 $\mu$m \citep{Turnbull06}
and the IRTF spectral library \citep{Rayner08}, 
which includes four planets (Jupiter, Saturn, Uranus and Neptune) 
taken with SpeX in the near-infrared (0.8-5.5$\mu$m)     

Most data for the spectral libraries were taken using a unique combination of telescope 
and instrument.  
The resolution, responses, and wavelength coverage were different for each data set.  
Moreover, it would be difficult to combine published spectral data in various wavelengths, 
because the data were taken in different phase or rotation of the planets. 
Few data, which were simultaneously obtained in a wide wavelength range from optical to infrared, 
are available. 

In the present study, we show the results of simultaneous spectroscopic observations in
wavelengths from optical to infrared for all planets, some moons, and some dwarf planets in
the solar system to study their general spectroscopic and photometric natures.
Accounts of the observations and data reduction are presented in \S 2 and \S 3, 
respectively.
The spectra of all targets we observed are compared in \S 4 along with the broad-band colors,
which are obtained by integrating the spectra.
The colors of other objects in the solar system, which are available in 
literature, are compared with our results. 
In \S 5, we discuss our results and relevant literature in conjunction with
the classification of gas, soil, and ice planets.

\section{Observations}

All observations for the present compilation of the spectral library 
were performed at Higashi-Hiroshima Observatory using the 1.5 m Kanata telescope and TRISPEC 
(Triple Range Imager and SPECtrometer with Polarimetry), 
a simultaneous optical and near-infrared imager, spectrograph, and polarimeter 
\citep{Watanabe05}.  
Two dichroic mirrors split the incoming light from the telescope into 
three beams, one optical channel (0.45 --0.90 $\mu$m) and two infrared channels IR1
(0.90 -- 1.85$\mu$m) and IR2 (1.85 -- 2.5 $\mu$m). 
It is capable of simultaneous three-band spectroscopy in a 
wavelength range of 0.45 -- 2.5 $\mu$m. 
We chose the slit of 300 $\mu$m width,
which gives the resolving power (R=$\lambda/\Delta\lambda$) of
R=138, 142, and 360 for optical, IR1 and IR2, respectively.
The TRISPEC slit lenght is 28 mm, which subtends 7 arcmin on the sky.

The observations were carried out during May 2-13, and November 19-27 of 2008.
While the overall data quality is good, shorter than 0.47 $\mu$m and between 0.9 and 1.0 $\mu$m 
the spectra are not reliable due to the dichroic coating problem with the beam splitters.
There is the possibility of thermal contamination from Earth's atmosphere and the 
telescope beyond 2.4 $\mu$m.
Around 1.4 and 1.8 $\mu$m, Earth's atmosphere has strong water absorption bands, where
the signal quality is low.

The bodies observed were: Mercury, Venus, Earth, Luna, Mars, Ceres, Jupiter, Io, 
Europa, Ganymede, Callisto, Saturn, Dione, 
Rhea, Titan, Uranus, Neptune, Pluto and Saturn's ring (Table 1).
We obtained Earth's spectra by observing earthshine, the light from the Earth which 
is reflected by the dark side of the moon.
We followed the procedure of Woolf et al. (2002) for acquiring and analyzing earthshine.
In May, we acquired a Lunar spectrum which corresponds to a bright crater,
while the bulk of the surface (mare) was observed in November.  

The log of the observations is listed in Table 1 along with brightness
in $V$ band, date, and phase angle (PhA) defined by the Sun: target: observer 
at the time of the observation.  
Earthshine shows us an integrated spectra of the entire Earth.  
Therefore, phase angle is the angle formed by the Sun: Earth: Luna.
Standard stars for the correction of atmospheric absorption and spectral 
response of TRISPEC were observed just before and after observation of each target.
The repeat observations give us information about sky background variation during 
observations. The standard star, its spectral type, and $V$
magnitude for each target are listed in the last three columns.

\begin{figure}
\resizebox{\hsize}{!}{\includegraphics{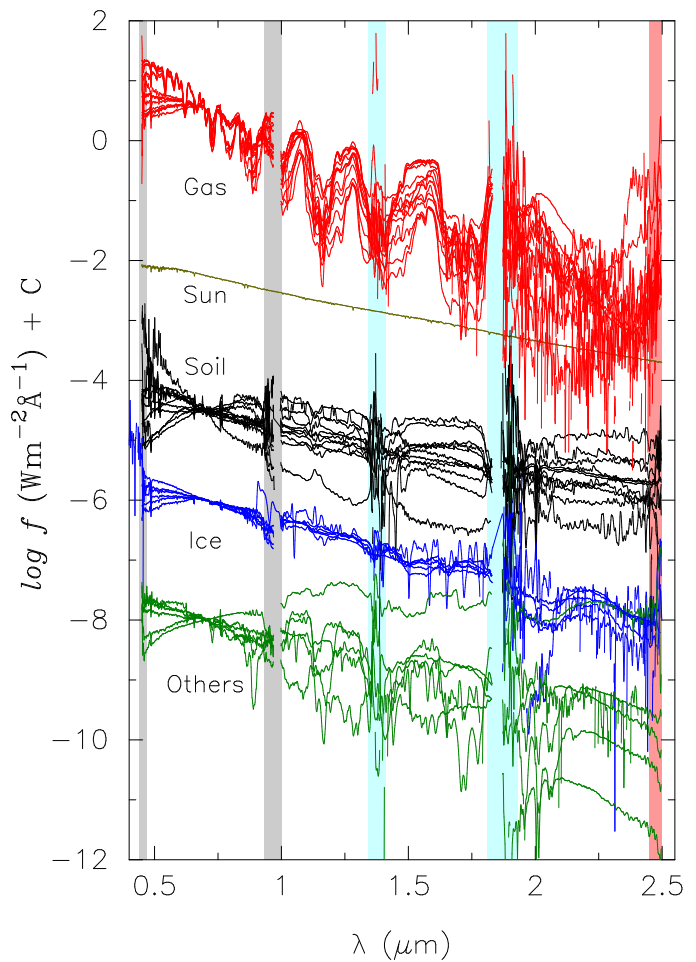}}
  \caption{All spectra of solar system objects taken by TRISPEC.
Each spectrum is normalized  at 0.7 $\mu$m. 
Flux, $f$, is arbitrarily offset so as to group the objects into gas, soil, ice, and anomalous others.
The Sun spectrum is taken from the STScI archive (http://www.stsci.edu/hst/observatory/cdbs/calspec.html).}
  \label{Fig1}
\end{figure}

\begin{table*}
\begin{center}

\caption{Observation Log}
\begin{tabular}{lcccccc}
\hline\hline
\noalign {\vspace{0.15cm}}
Target & Date & Phase & $V$  & Standard & Spectral& $V$ \\
 & & Angle & & Star{$^\dagger$} & Type &  \\
 & & (deg) & (mag) & & & (mag) \\
\noalign {\vspace{0.15cm}}
\hline
\noalign {\vspace{0.15cm}}
May 2008 \\
\cline{1-1}
\noalign {\vspace{0.15cm}}
Callisto  & 05/05/08 & 10    &  6.3 & BS7264 & F2II & 2.89 \\
Saturn    & 05/05/08 & 6     &  1.0 & BS4101 & A0 & 2.89 \\
Dione     & 05/05/08 & 6     & 10.4 & BS4101 & A0 & 2.89 \\
Titan     & 05/(05,06)/08 &  6   & 8.4 & BS4101 & A0 & 2.89 \\
Io        & 05/06/08 & 10    &  5.4 & BS7264 & F2II & 2.89 \\
Saturn's Ring      & 05/06/08 & 6     &  1.0 & BS4101 & A0 & 2.89 \\
Rhea      & 05/06/08 & 6     &  9.7 & BS4101 & A0 & 2.89 \\
Jupiter   & 05/08/08 & 10    & -2.4 & BS7264 & F2II & 2.89 \\
Europa    & 05/07/08 & 10    &  5.6 & BS7264 & F2II & 2.89 \\
Neptune   & 05/07/08 & 2     &  7.9 & HIP105946 & B9III/IV & 7.09 \\
Uranus    & 05/(07,11)/08 & 2   & 5.9 & BS8911 & A0p & 4.49 \\
Mercury   & 05/11/08 & 96    &  0.0 & BS1641 & B3V & 3.17 \\
Earth     & 05/11/08 & 81,84 & -2.5 & BS2890 & A2V & 1.58 \\
Luna (crater) & 05/11/08 & 97 & -9.8 & BS2890 & A2V & 1.58 \\
Pluto     & 05/11/08 & 1     &  15  & HIP87881  & A0V & 10.08 \\
Mars      & 05/12/08 & 35    &  1.3 & BS2890 & A2V & 1.58 \\
\noalign {\vspace{0.15cm}}
\cline{1-1}
\noalign {\vspace{0.15cm}}
Nov 2008 \\
\noalign {\vspace{0.15cm}}
\cline{1-1}
\noalign {\vspace{0.15cm}}
Saturn's Ring      & 11/19/08 & 6 & 1.2              & BS4101    & A0 & 2.89 \\
Saturn    & 11/(19,22)/08 & 6 & 1.2      & BS4101    & A0 & 2.89 \\
Neptune   & 11/(20,25,26)/08 & 2 & 7.9   & HIP104167 & A0 & 10.10 \\
Venus     & 11/(20,26)/08 & 63,65 & -4.2 & BS7264    & F2II & 2.89 \\
Uranus    & 11/20/08 & 3 & 5.8              & HIP112880 & A0 & 8.22	\\
Earth     & 11/21/08 & 70 & -2.5         & HIP50793  & A2 & 9.13 \\
Luna (mare) & 11/21/08 & 108,109 & -9.3       & BS4101    & A0 & 2.89 \\
Titan     & 11/25/08 & 6 & 8.6              & HIP50793  & A2 & 9.13 \\
Ceres     & 11/25/08 & 23 & 8.4             & HIP50793  & A0 & 9.13 \\
Rhea      & 11/25/08 & 6 & 9.9              & HIP50793  & A2 & 9.13 \\
Dione     & 11/25/08 & 6 & 10.6             & HIP50793 & A2 & 9.13 \\
Io        & 11/26/08 & 8 & 5.8              & HIP106831 & B9V & 8.27 \\
Ganymede  & 11/26/08 & 8 & 5.4              & HIP106831 & B9V & 8.27 \\
\noalign {\vspace{0.20cm}}
\hline
\noalign {\vspace{0.20cm}}
\multicolumn{7}{l}{\footnotesize
$^\dagger$ BS and HIP refer to UKIRT (2008) and Perryman et al. (1997), respectively.
}
\end{tabular}
\label{table:1}
\end{center}
\end{table*}

Special care was taken in the observations for large objects, bright objects, Saturn 
and its ring.

Large targets: 
The angular sizes of Jupiter and Saturn are larger than the 4.5 arcsecond angle subtended by the 
spectrograph slit.
To take spectra of these planets, we aligned the slit along a north/south line through 
the center of planet; each planet's axis of rotation is at a slight angle to the slit.  
These spectra do not include light from the entire planet as the spectrum for an
object smaller than the slit width would;
only the part visible through the slit is sampled.
We will use these spectra as first approximations for the reflection spectra of unresolved gas giants,
and note that the scatter in the multiple observations of Saturn would probably decrease 
if we were able to sample more of the surface.
The spectra for sky subtraction were obtained from the sky adjacent to the target.

For earthshine observations, the lunar disk fills the entire spectrograph slit, 
so that we had to take dedicated sky background frames to the East and West of Luna.  
These frames were taken using the same exposure times and procedures as observations of the dark portions of Luna.  
The illuminated portion of Luna is bright enough that sky glow does not contribute noticeably, 
and can be safely neglected \citep{Turnbull06}.  
We took sky background frames to the East and West.  
Our background frames were placed 30 arc minutes in front of and behind Luna's center point.  
Our earthshine observation was made at 10 arc minutes from the center on the dark side and 
our observations for Luna was made at 10 arc minutes on the illuminated side.  
Thus we had equal spacing of 20 arc minutes between the exposure points. 
This allowed us to account for light from the bright surface which had been scattered 20 degrees and 
would thus show up in our dark side exposure.

Bright targets: 
For bright targets such as Jupiter and crescent Luna, detector saturation 
became a problem.
We partially closed the telescope dome slit 
to reduce the light gathering power of the telescope.
We used this technique only with bright objects, so that 
thermal noise from the dome did not have a noticeable effect on the signal quality
even near 2.5 $\mu$m.

Saturn and its ring: 
Because of the north/south orientation of the slit, most of the signals shown 
for Saturn include contamination from the planet's strong ring system.  
We expect that the ring spectrum and gas planets spectrum do not resemble each other.  
Therefore, we obtained a spectrum for Saturn without its ring during the November observations,
by rotating the spectrograph slit 90 degrees and placing it on Saturn's north pole.
As stated above, we are aware that sampling a small part of the surface does not necessarily
reflect the integrated spectrum of the planet, but we will use it as an approximation of such.

\section{Data Analysis}

We followed the standard procedure for flat fielding, sky subtraction and spectrum extraction,
using IRAF.
To correct for sky absorption and the spectral response of TRISPEC, 
we observed standard stars with airmasses similar to the target.
Because the exposure time used for each of TRISPEC's three channels has an effect 
on the actual exposure time of the other two channels we made sure to use exactly the same exposure times
for the standard as for the target.
After the spectra had been extracted and wavelength calibrated, 
we divided each planet spectrum by observed spectrum of the corresponding standard star
and multiplied by the calculated actual spectrum for the star's spectral type 
taken from the ISAAC database of simulated spectral-type spectra \citep{Pickles98}.
The standard stars we used for each object and their spectral types are listed in Table 1.

For moons, there is the special problem of light from the parent planet.
The amount of scattered light occuring in the telluric atmosphere or in the telescope
is a function of radius from the planet.  
Thus lines of equal light pollution form rings around the planet.
During data analysis we learned the hard way that, when we observe a moon close to the parent planet, 
the straight line of the spectrograph slit does not offer a good approximation 
to the rings of equal light pollution.
Even though we subtracted the adjacent sky signal, 
it was not sufficient to completely remove the scattered light from the planet.
When observing moons, we should align the spectrograph slit along the moon's orbit plane.
This way the sky used for sky subtraction will include areas 
which have more and less light pollution than the sky in front of the target, 
so that we can take an average to better determine the actual amount of scattered light 
in the telluric atmosphere infront of the target.

In analyzing spectra of the solar system planets, 
authors studying a particular planet's reflectivity often divide the planet spectra 
by the Sun's spectrum.
Groups focusing on a single feature will normalize to a continuum instead.
Because there is no consensus on which method to use, 
we use the spectra without applying either method for the following discussion.
The earthshine data requires more procedures than the other observations.
Because the light we see as earthshine is reflected by Luna, the spectrum is affected by the wavelength dependence of Luna's reflectivity.
Therefore, we divided the earthshine spectra by a spectrum of moonlight obtained on the same night, 
with corrections for phase angle \citep{Woolf02}.
Most earthshine research groups publish their results in terms of Earth's reflectivity;
we have taken the additional step of multiplying by the Sun's spectrum 
to compare Earth's spectrum to the rest of the planets.

\section{Results}
All spectra we analyzed are shown in Fig. \ref{Fig1} and Appendix A.
The spectra are normalized at 0.7 $\mu$m.
In this paper, we use the term 'planet' loosely, to include moons and rings as well.  
Several objects were observed multiple times as listed in Table 1.
These multiple observations are plotted separately in Fig. \ref{Fig1}, 
because spectra depend on the object's rotation, 
and changes in phase angle or the slit location on extended objects.
For example the spectrum of Mars shows dramatic changes based on the part of the surface visible;
and individual sections of Mars show time dependent changes \citep{McCord71}.
We also show the spectrum of the Sun for comparison.

\subsection{Classification by Spectral Features}
To stress the patterns which emerge by comparing the spectra,
we classify the planets into gas, soil, and ice from their spectral characteristics with
the following criteria. 
There are also some anomalous spectra which cannot be easily attributed to one of the categories.  

Gas Planets -- Jupiter, Saturn, Uranus, Neptune, Titan:
The spectra of gas planets are dominated by strong methane absorption around 1.0, 1.17, 1.7, 
and 2.2 $\mu$m \citep{Strong93}.  
We specifically avoid the term 'gas giant' because Titan's spectra match the others so well.  
The atmospheres of Jupiter, Saturn, Uranus and Neptune are composed primarily of hydrogen and helium.
The bulk of Titan's atmosphere is nitrogen.  
Small amounts of methane are enough to cause deep, broad absorption features.
Changes in the shapes of these features indicate the concentration of methane.

Soil Planets -- Mercury, Earth, Luna, Mars, Ceres, Io, Ganymede:
Mars is a typical sample of the category, which we call soil planet.
Compared to the gas planets, soil planets have a simple shape.  
The flux remains roughly level from the optical through near infrared.
One exception is the Luna crater, which shows a steep spectrum in visible (see \S 5 for more details).  

Ice Planets -- Saturn's Ring, Dione, Europa, Rhea, Pluto:
The group is characterized by broad and deep absorption of crystalline water ice centered 
near 2.05 $\mu$m, roughly 0.3 $\mu$m wide. 
The steep blue slope and absorption lines of ice planets match the spectra of water ice in the laboratory \citep{Clark81,Grundy99}.  
Pluto's spectrum has a slight wave caused by its methane ice composition \citep{Protopapa08}, 
but the overall shape is very similar to water ice.

Others:
We place the four objects with contaminated spectra into the Others class of objects.
They are the May observations for Callisto, Io, Rhea and the November observation of Dione.  
For all objects except Rhea, we were also able to obtain an uncontaminated spectrum.

\subsection{Color-Color Distribution}
\begin{figure}
\resizebox{\hsize}{!}{\includegraphics{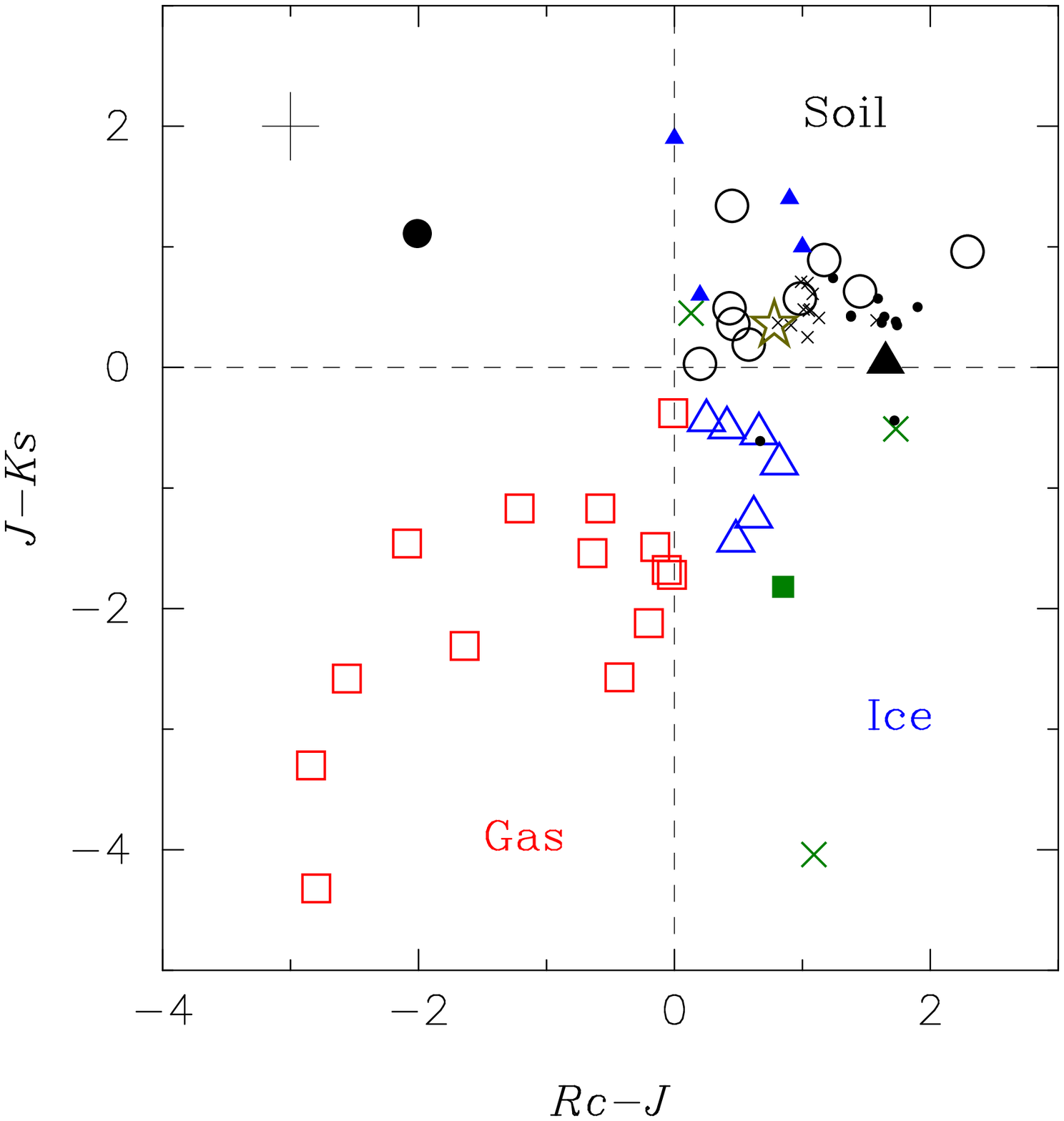}}
  \caption{Color-color diagram for the objects listed in Table \ref{table:2}.  
The magnitudes are obtained by integrating the spectra. 
Circle, triangle, square, and cross show the classes of soil, ice, gas, and anomalous others 
(see Table \ref{table:2}).  
A filled circle, filled squares and a filled large triangle 
are Luna crater, Venus, and Ganymede, respectively.
Small dots show Kuiper belt objects \citep{Delsanti04}.  
Small crosses are trans neptunian objects and centaurs \citep{Demeo09}. 
Filled small triangles are comet Halley 1986 III (Gehrz \& Ney 1992). 
The solar color is shown by a star. 
The typical error bars of our observations are depicted at the upper left
corner.
}
\label{Fig2}
\end{figure}

One inherent disadvantage of spectroscopy is that it requires long observation times.
For dim objects and surveys of a large number of objects color observations are more convenient.  
From our spectral data, we calculate the colors of the planets.  
The magnitudes are defined by the flux integrated with filter responses 
of $V$ ($\lambda_{\mathrm eff} = 0.55$ $\mu$m) \citep{Johnson53}, $Rc$ 
(0.66 $\mu$m), $Ic$ (0.81 $\mu$m) \citep{Cousins78}, and $J$ (1.25 $\mu$m), 
$H$ (1.64 $\mu$m), $Ks$ (2.15 $\mu$m) \citep{Tokunaga02}.  
Colors are defined so as to be 0.0 for all colors of Vega, of which the
spectrum is taken from http://www.stsci.edu/hst/observatory/cdbs/calspec.html. 
The colors of Jupiter, Saturn, and Neptune taken with SpeX (Rayner et al. 2008) were also 
obtained in the same manner.
The results of the color analysis are shown in Table \ref{table:2}. 
The repeated observations of the standard stars before and after a
target allow us to estimate the errors.
The resultant errors in optical and infrared colors are
$\Delta(V-Rc)=0.13$, $\Delta(R-Ic)=0.09$, $\Delta(Rc-J)=0.22$,
 $\Delta(J-Ks)=0.28$, and $\Delta(H-Ks)=0.30$.
The errors mainly originated from variable sky background during a target's observations.
The larger errors in infrared would be due to variation of water vapor in the telluric atmosphere.
In Fig. \ref{Fig2} we plot the $J-K$ vs. $Rc-J$ colors.
Again, in Fig. \ref{Fig2} and Table \ref{table:2}, multiple observations of the same target are shown.
Some of the scatter is the result of photometeric error, but 
we cannot rule out the possibility that there are also time dependent changes in the target.
For example, an extended photometric time series in $J$ and $K$ bands
of Neptune showed rapid color variation in a day by $\sim3$
mag in $(J-K)$ due to the planet's rotation \citep{Belton81}.

We plot additional color information for Kuiper belt objects (KBO) \citep{Delsanti04}, 
trans neptunian objects (TNO), centaurs \citep{Demeo09}, and comet Halley 1986 III 
\citep{Gehrz92}, for which $Rc$, $J$, and $K$  data are available.

\begin{table*}
\begin{center}
\caption{Photometric Colors of Targets}
\label{table:2}
\begin{tabular}{lcrrrrrl}
\hline\hline
\noalign {\vspace{0.15cm}}
 Object                & Class &  \multicolumn{1}{c}{$V-R$}   & \multicolumn{1}{c}{$R-Ic$} & \multicolumn{1}{c}{$R-J$} & \multicolumn{1}{c}{$J-K$} & \multicolumn{1}{c}{$H-Ks$}    &   Remark           \\
\hline
\noalign {\vspace{0.15cm}}
 The Sun               &       &    0.38 & 0.35 & 0.78  & 0.35 &  0.05  &   (1)             \\
\cline{1-1}
\noalign {\vspace{0.15cm}}
May 2008 \\
\cline{1-1}
\noalign {\vspace{0.15cm}}
 Earth (PhA=81$^\circ$) & Soil &   0.09 &   0.02 &   0.43 &   0.49 &   0.56 &   desert $\&$ ocean \\ 
 Earth (PhA=84$^\circ$) & Soil &   0.13 &   0.15 &   0.20 &   0.03 &   0.21 &   desert $\&$ ocean \\
 Luna (crater)          & Soil &  -0.88 &  -0.81 &  -2.01 &   1.11 &   1.27 &  descending (2)      \\
 Mercury                & Soil &   1.17 &   1.17 &   2.29 &   0.96 &   0.37 &                     \\
 Mars                   & Soil &   1.03 &   0.90 &   1.17 &   0.89 &   0.37 &                     \\
 Pluto                  & Ice  &   0.35 &   0.28 &   0.41 &  -0.50 &  -0.82 &                     \\
 Dione                  & Ice  &   0.12 &   0.08 &   0.25 &  -0.44 &  -0.23 &                     \\
 Europa                 & Ice  &   0.69 &   0.49 &   0.82 &  -0.80 &  -0.53 &                     \\
Saturn's Ring           & Ice  &   0.46 &   0.41 &   0.62 &  -1.24 &  -0.89 &  (2)                \\
 Jupiter                & Gas  &   0.28 &   0.06 &  -0.64 &  -1.54 &  -1.33 &  (2)                \\ 
 Saturn                 & Gas  &   0.18 &  -0.12 &  -0.20 &  -2.12 &  -2.31 &   w/ ring (2)       \\
 Titan (05/05)          & Gas  &   0.37 &   0.01 &  -0.58 &  -1.17 &  -0.30 &                     \\
 Titan (05/06)          & Gas  &   0.67 &   0.33 &  -0.01 &  -0.38 &  -0.10 &                     \\
 Uranus (05/11)         & Gas  &  -0.28 &  -1.01 &  -2.84 &  -3.30 &  -3.31 &                     \\
 Neptune                & Gas  &  -0.17 &  -0.78 &  -1.64 &  -2.31 &  -2.14 &                     \\
 Callisto               & --   &   0.25 &  -0.17 &   1.09 &  -4.04 &  -3.95 &                     \\
 Io                     & --   &   0.81 &   0.36 &   1.73 &  -0.51 &  -0.26 &                     \\
 Rhea                   & --   &   0.98 &   1.04 &   3.68 &   1.09 &   0.76 &                     \\
\noalign {\vspace{0.15cm}}                                           
\cline{1-1}
\noalign {\vspace{0.15cm}}
Nov 2008 \\
\noalign {\vspace{0.15cm}}
\cline{1-1}
\noalign {\vspace{0.15cm}}
 Earth                        & Soil &   0.08 &   0.00 &   0.58 &   0.19 &  -0.06 &      ocean    \\ 
Luna (mare) (PhA=108$^\circ$) & Soil &   0.39 &   0.43 &   0.98 &   0.57 &   0.03 &  ascending (2)\\
Luna (mare) (PhA=110$^\circ$) & Soil &   0.52 &   0.32 &   0.45 &   1.34 &   0.81 &  ascending (2) \\
 Ceres                        & Soil &   0.19 &   0.24 &   0.46 &   0.36 &   0.06 &              \\
 Ganymede                     & Soil &   1.04 &   0.87 &   1.65 &   0.04 &  -0.08 &              \\
 Io                           & Soil &   0.62 &   0.71 &   1.45 &   0.63 &  -0.10 &              \\
 Rhea                         & Ice  &   0.29 &   0.17 &   0.48 &  -1.44 &  -1.26 &              \\
 Ring                         & Ice  &   0.54 &   0.31 &   0.66 &  -0.55 &  -0.43 &   (2)        \\
 Saturn (11/19)               & Gas  &   0.48 &   0.12 &  -0.06 &  -1.68 &  -1.79 &   w/ ring (2)\\
 Saturn((11/19)               & Gas  &   0.54 &   0.13 &  -0.02 &  -1.72 &  -1.85 &   w/o ring, N.pole(2)   \\
 Saturn (11/22)               & Gas  &   0.26 &   0.03 &  -0.15 &  -1.49 &  -1.49 &   w/ ring (2)\\
 Titan                        & Gas  &   0.55 &   0.06 &  -0.43 &  -2.57 &  -2.20 &              \\
 Uranus                       & Gas  &  -0.25 &  -0.87 &  -2.80 &  -4.32 &  -4.20 &              \\
 Neptune  (11/20)             & Gas  &  -0.32 &  -0.81 &  -2.56 &  -2.58 &  -2.55 &              \\
 Neptune  (11/25)             & Gas  &  -0.32 &  -0.78 &  -2.09 &  -1.46 &  -1.53 &              \\
 Neptune  (11/26)             & Gas  &  -0.12 &  -0.73 &  -1.21 &  -1.17 &  -1.14 &              \\
 Venus    (11/20)             & --   &   0.38 &   0.40 &   0.85 &  -1.82 &  -1.97 &   (3) \\
 Venus    (11/26)             & --   &   0.18 &   0.02 &  -1.70: &   2.40: &   1.61: &   (3) \\
 Dione                        & --   &   0.18 &   0.12 &   0.13 &   0.45 &   0.23 &           \\
\noalign {\vspace{0.15cm}}
\cline{1-1}
\noalign {\vspace{0.15cm}}
SpeX (Rayner et al. 2008) \\
\noalign {\vspace{0.15cm}}
\cline{1-1}
\noalign {\vspace{0.15cm}}
Jupiter               & Gas  & & & & -2.75 & -2.55 &  \\
Saturn                & Gas & & & & -1.78  & -1.96  & \\
Neptune               & Gas & & & & -2.63 &  -2.58 & \\
\noalign {\vspace{0.20cm}}
\hline
\noalign {\vspace{0.20cm}}
\multicolumn{8}{l}{\footnotesize
(1) The flux data is taken from http://www.stsci.edu/hst/observatory/cdbs/calspec.html. 
}\\
\multicolumn{8}{l}{\footnotesize
(2) Because the target is larger than the spectrograph slit,these spectra are from part of the
}\\
 \multicolumn{8}{l}{\footnotesize
surface instead of the integrated planet spectrum.
}\\
\multicolumn{8}{l}{\footnotesize
(3)  The signal in November 20 for Venus is in good agreement with the spectra taken by Pollack (1978).
}\\
\multicolumn{8}{l}{\footnotesize
Compared to the Nov 20th spectrum, the Nov 26th signal has the same shape, but the region from
}\\
\multicolumn{8}{l}{\footnotesize
1 to 1.8 $\mu$m appears depressed, though the origin is not known.
}\\
\end{tabular}
\end{center}
\end{table*}                                               

\section{Discussion}

In the following discussion we use the word 'blue' to mean that 
the signal is stronger at shorter wavelengths than at longer wavelengths.
Conversely, 'red' is used to mean the signal is stronger at longer wavelengths.

The overall slope of the gas planet spectra across the infrared is very blue.
The exact shapes of the spectra are determined by the composition of each planet's atmosphere and clouds.
A full discussion of the scattering mechanisms of these 5 objects is beyond the scope of this paper.
For our purposes it is sufficient to note that slope can be modeled 
with a simple  $\lambda^-4$ Rayleigh scattering model \citep{Bergstralh83}.

The spectra of the gas planets diverge in the visible.  
In the visible Uranus and Neptune are very blue, with strong methane absorption lines, while Jupiter, Saturn and Titan have level spectra.   
Jupiter and Saturn have high clouds which are more reflective in the visible than in the 
infrared\citep{Fink79}.   
The reflection/transmission properties of the clouds on Uranus and Neptune are less dependent on wavelength,
so that short wavelength light exhibits the same steep blue trend and strong methane absorption as we 
see in the infrared \citep{Fink79,Karkoschka94}.  
On Titan, aerosols prevent visible light from reaching the same depths in 
the atmosphere as infrared light \citep{Fink79,Karkoschka94}. 

Soil planets also have a large dispersion in the optical spectra.  
This is not surprising, given the vast differences in the surfaces of Mercury, Earth, and Ceres.
Individual spectral features come primarily from silicates, iron bearing minerals, 
and water in the form of hydrated minerals or ice.
The Luna crater is the bluest spectrum in the soil category of Fig. \ref{Fig1}.  
This spectrum is similar to infrared crater spectra 
obtained by other researchers \citep{McCord81}.
The spectrum  drops sharply below the other soil planets in the visible and stays 
flat though the infrared.  
Bright lunar craters contain freshly exposed or young material.  
The gradual reddening (i.e. the decrease of reflectance at short wavelengths) 
is a result of the weathering process of agglutinate accumulation on the surface 
\citep{Pieters86}.

In contrast to the spectra of gas and soil planets, 
the spectra for ice planets varies little with wavelength.
While the other two categories have a large dispersion, 
the individual ice planet spectra closely resemble one another. 
Even though Pluto is primarily methane ice instead of water ice  \citep{Protopapa08}
it is still a close match for the other ice planets.

Although many photometric observations were performed so far, most were observed in different 
epoch in optical and infrared. 
Therefore our color-color diagram (Fig. 2) would reflect a more real nature of the planets 
than previous studies. 
On the color-color diagram the differences between planet types suggested by the individual 
spectra become obvious.
We roughly divide the figure into Gas, Soil, and Ice regions, though some targets are out of the categories.

Soil planets are close to the Sun's color.  
Ice planets are bluer in $J-K$ than soil planets; gas planets are much bluer in both $Rc-J$ and $J-K$.  
Although $Rc$-band data is not available, the colors of asteroid in infrared \citep{Baudrand01, 
Baudrand04}, which are distributed in $J-K> 0$ and $I-J>0$, are consistent with the soil class.
The location of a blue KBO, (24835) 1995 SM-55, ($Rc-J=0.67$, $J-K=-0.61$) 
supports the finding indicative of water ice absorption features \citep{Delsanti04}.

It is interesting to note that even though the Others class of objects were contaminated by 
stray light from their parent gas planet, on the color-color diagram
 none of them fall into the gas planet area.
The contamination was not strong enough to change the color seriously.

There are two topics which deserve special attention:  Ganymede \& dirty ice and Venus. 

Ganymede \& Dirty Ice: 
We identified Ganymede as soil because its spectrum has more 
in common with Mars and Ceres than with the spectra of icy satellites. 
While Ganymede has a rocky core, the surface is primarily ice. 
Estimates of the amount of ice on Ganymede's surface have been revised over the years.  
Current models take into account dirty ice contaminated with dark particles, probably 
an iron rich material \citep{Clark80}.  
The dark particles can be dredged up from Ganymede's large rocky core by ice volcanism \citep{Showman04}.
This justifies our classification of Ganymede as soil 
and shows that there is not a sharp division between soil planets and ice planets.  
There is a gradual transition through dirty ice as the fractional composition changes.  

Comets are well known to be 'dirty snowballs.'  
Colors are shown for Halley's comet \citep{Gehrz92} in Fig. \ref{Fig2} .
The colors fall in the range of the soil planets, and appear to be slightly redder than soil 
planets in $J-Ks$ color.  
Laboratory work with mineral grains on water ice frost shows that small amounts of silicate 
contamination can produce a dramatic change in the near infrared spectra \citep{Clark81}.  
In particular there is a dramatic reddening in the wavelengths used to calculate $J-K$ color 
\citep{Clark80}.

Venus:
Venus does not fit to the soil planet spectra.  
This is not unexpected since the solid surface of Venus is hidden beneath a thick 
cloud bank. 
Venus has a $(J-K)$ color which is bluer than any of the soil planets.
This color is caused by scattering in the Venus's thick atmosphere.
Venus does match the spectra for ice planets fairly closely (see appendix A).

Distinguishing between Venus and ice planets requires being able to 
differentiate between H$_2$0 and CO$_2$ absorption features.
Venus shows strong CO$_2$ lines at 1.4, 1.6 and 2.0 $\mu$m. 
Water ice has strong lines at 1.5, 1.65 and 2.0 $\mu$m.
The ice lines are known to shift based on the physical structure and temperature of the 
ice \citep{Grundy99}, making distinguishing between the two more difficult.  
The two can be distinguished by their spectra but not by their color.
More work, possibly in the visible or ultraviolet, is needed to break the Venus/ice degeneracy. 	
It is possible that Venus represents a fourth class, such as veiled planets, 
which have a thick atmosphere devoid of the methane features found in gas planets.

\section{Conclusion}

We obtained spectra for all of the solar system planets and many smaller bodies
using TRISPEC on the 1.5m  Kanata telescope at Higashi-Hiroshima Observatory. 
The spectra are divided by their characteristic shapes into three distinct groups: gas planets, soil planets, and ice planets.  
Venus represents a possible fourth group. 
Our results indicate that low resolution, broad-band spectroscopy across the visible and near infrared can provide a wealth of information about the planets being observed.  

In addition to spectroscopy, simple color observations can be used to separate planets into these three categories.  
Although Venus and Ganymede show that the classification system isn't perfect, 
color observations have the advantage that they can be used on fainter and/or more numerous objects than even low 
resolution spectroscopy.

We originally undertook this research towards understanding extrasolar planets but
even within our own solar system, there are numerous small bodies whose compositions 
remain unidentified, or implied through density calculations which require difficult to 
obtain observations \citep{Grundy07}.  
Our work demonstrates the potential for meaningful data in visible and near infrared color surveys 
of a large number of asteroids, centaurs, and TNO.
Our results will provide useful comparison for reflection spectra for exoplanets.

The Tohoku-Hiroshima-Nagoya Planet Spectra Library (THN-PSL) presented here is 
the first library to contain visible and near IR spectra for all of the solar system planets
observed with same instrument and telescope combination for all planets. 
This library will be a useful reference for analyzing extrasolar planet spectra, and for 
comparing between planetary data sets.

\begin{acknowledgements}
We thank all members of the TRISPEC development team and the Higashi-Hiroshima Observation team 
for their hard work which helped to make our observations a success.  
This work was partially supported by the Mitsubishi Foundation.
IRAF is distributed by the National Optical Astronomy Observatories, which are operated by the Association of Universities for Research in Astronomy, Inc., under cooperative agreement with the National
Science Foundation.  
\end{acknowledgements}

\begin{appendix}
\section{Complete Spectral Data}
In this appendix we present all of our observed spectra.
The spectra are normalized at 0.7  $\mu$m.
The text data are available upon request to the authors (RL, TI).

\begin{figure*}
\resizebox{\hsize}{!}{\includegraphics{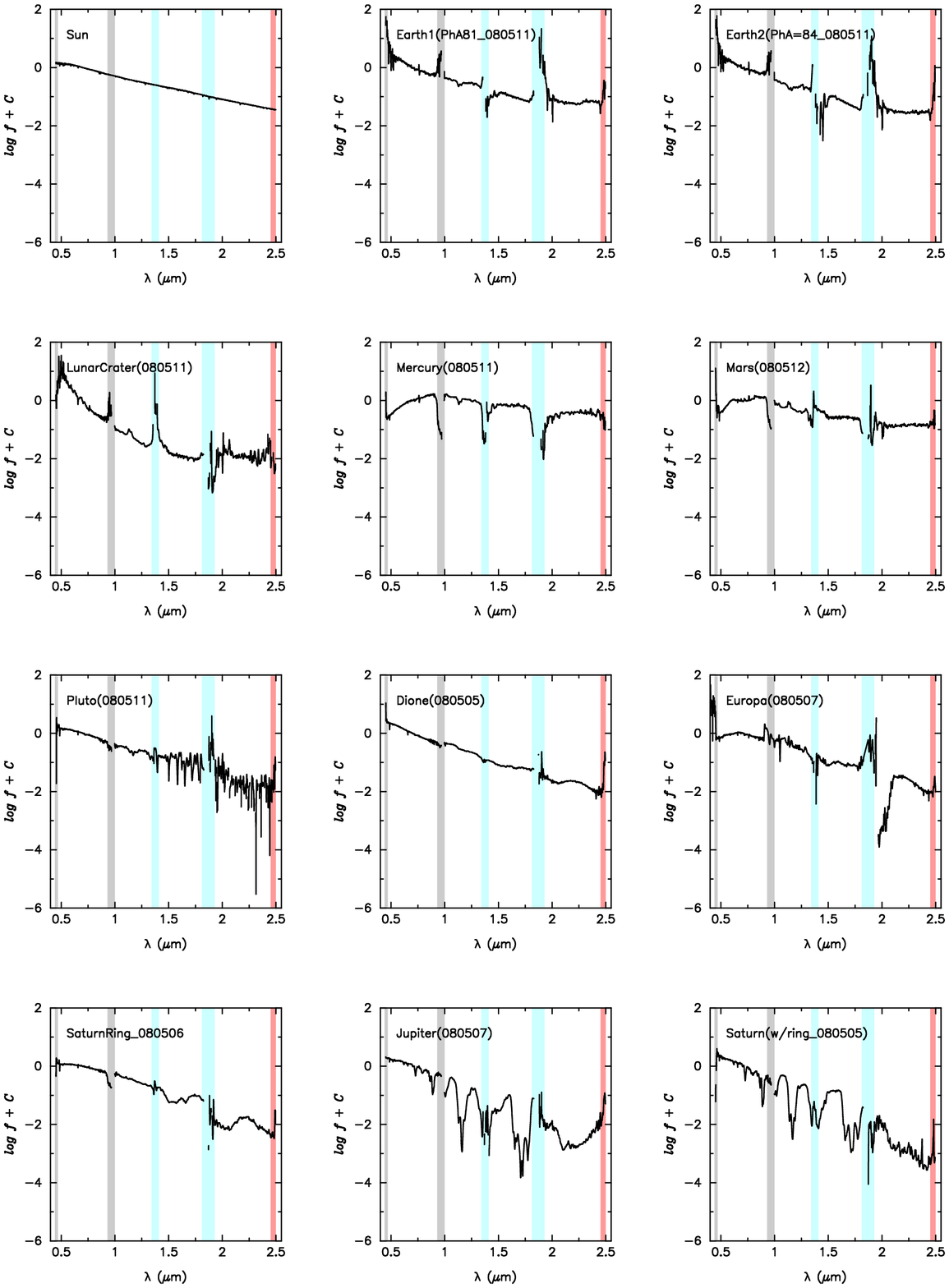}}
\label{A1}
\end{figure*}

\begin{figure*}
\resizebox{\hsize}{!}{\includegraphics{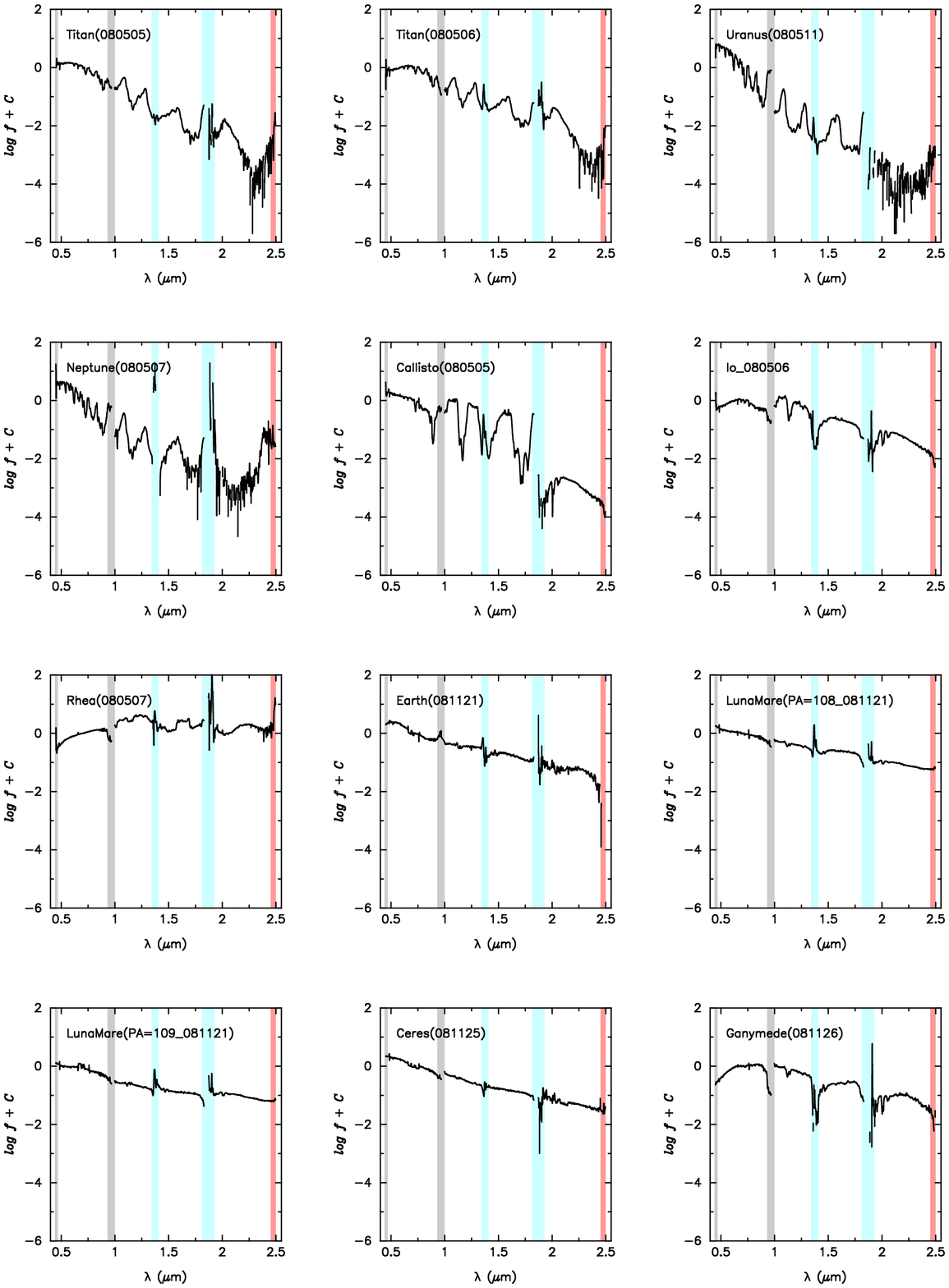}}
\label{A2}
\end{figure*}

\begin{figure*}
\resizebox{\hsize}{!}{\includegraphics{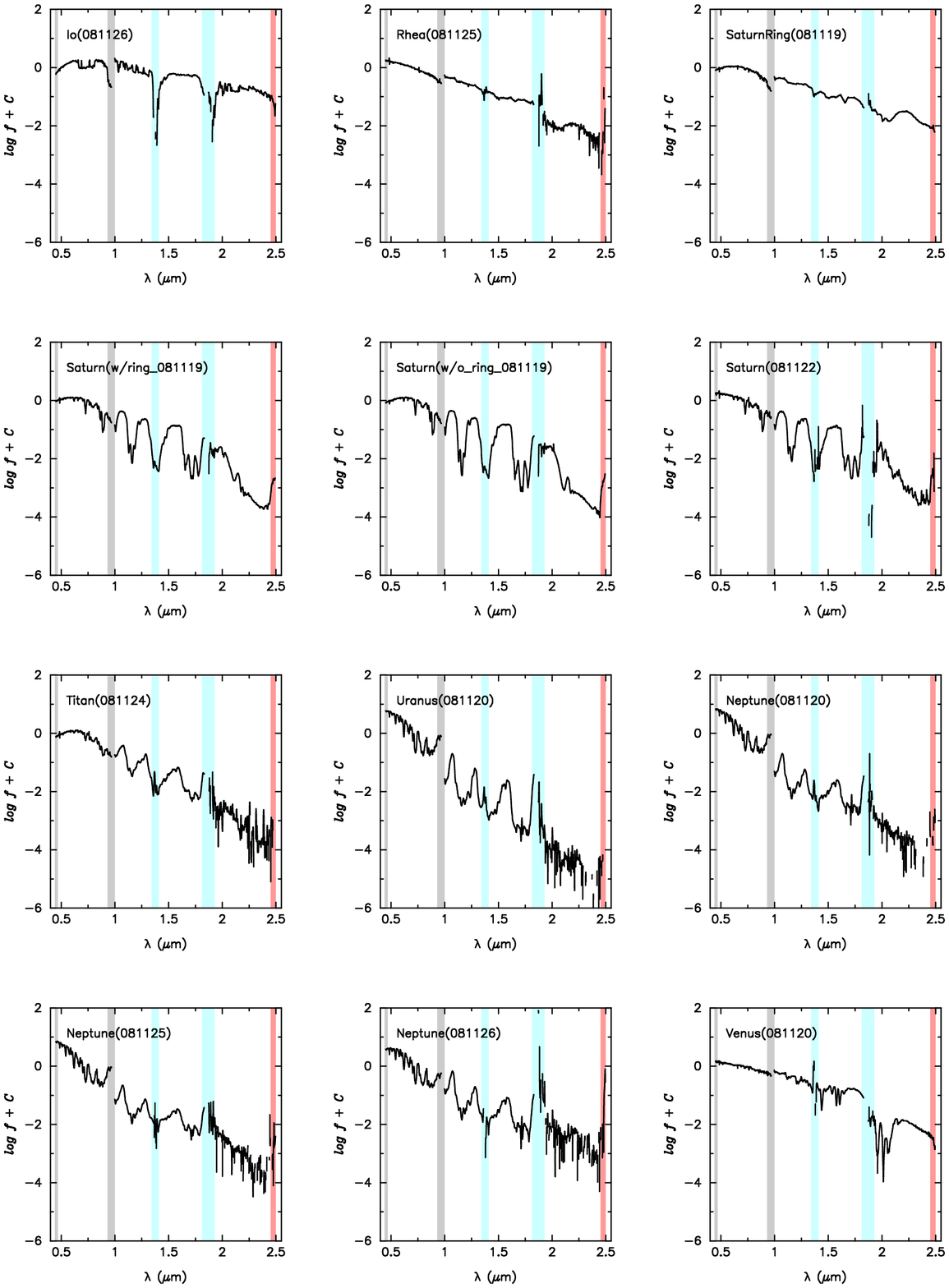}}
\label{A3}
\end{figure*}

\begin{figure*}
\resizebox{\hsize}{!}{\includegraphics{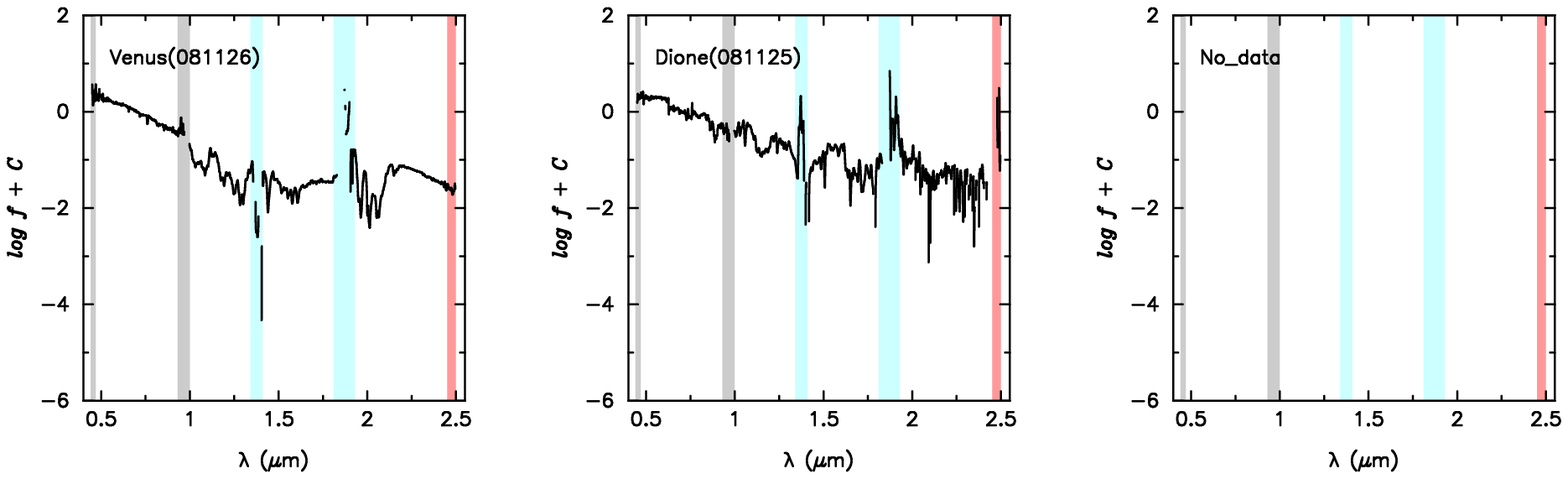}}
\label{A4}
\end{figure*}

\end{appendix}

\bibliographystyle{aa}
\end{document}